\documentclass{svproc}
\usepackage{url}

\usepackage{graphicx}
\usepackage{amsmath}
\usepackage{amsfonts}
\usepackage{amssymb}
\usepackage{bm}
\usepackage{braket}
\usepackage{subfigure}
\usepackage{hyperref}
\usepackage{cite}
\usepackage{color}
\usepackage[normalem]{ulem}

\newcommand{\be}{\begin{equation}}
\newcommand{\bee}{\begin{equation*}}
\newcommand{\ee}{\end{equation}}
\newcommand{\eee}{\end{equation*}}
\newcommand{\bearre}{\begin{eqnarray*}}
\newcommand{\eearre}{\end{eqnarray*}}
\newcommand{\bearr}{\begin{eqnarray}}
\newcommand{\eearr}{\end{eqnarray}}

\begin{document}
\mainmatter              
\title{Characterising two-sided quantum correlations beyond entanglement via metric-adjusted $f-$correlations}
\titlerunning{Quantum metric-adjusted $f-$correlations}  
%
\author{Marco Cianciaruso\inst{1} \and Ir\'{e}n\'{e}e Fr\'{e}rot\inst{2} \and
Tommaso Tufarelli\inst{1} \and Gerardo Adesso\inst{1}}
\authorrunning{Marco Cianciaruso et al.} 
%
%
\institute{Centre for the Mathematics and Theoretical Physics of Quantum Non-Equilibrium Systems, School of Mathematical Sciences, The University of Nottingham, University Park, Nottingham NG7 2RD, United Kingdom\\
\email{gerardo.adesso@nottingham.ac.uk}
\and
Univ Lyon, Ens de Lyon, Univ Claude Bernard, CNRS, Laboratoire de Physique, F-69342 Lyon, France}

\maketitle              

\begin{abstract}

We introduce an infinite family of quantifiers of quantum correlations beyond entanglement which vanish on both classical-quantum and quantum-classical states and are in one-to-one correspondence with the metric-adjusted skew informations. The `quantum $f-$correlations' are defined as the maximum metric-adjusted $f-$correlations between pairs of local observables with the same fixed equispaced spectrum. We show that these quantifiers are entanglement monotones when restricted to pure states of qubit-qudit systems. We also evaluate the quantum $f-$correlations in closed form for two-qubit systems and discuss their behaviour under local commutativity preserving channels. We finally provide a physical interpretation for the quantifier corresponding to the average of the Wigner-Yanase-Dyson skew informations.

\end{abstract}
\section{Introduction}\label{sec:introduction}
Nonclassical correlations in quantum systems manifest themselves in several forms such as non-locality~\cite{bell1966problem,brunner2014bell}, steering~\cite{wiseman2007steering,cavalcanti2016quantum}, entanglement~\cite{horodecki2009quantum}, and discord-type quantum correlations beyond entanglement~\cite{modi2012classical,streltsov2014quantum,adesso2016measures,fanchini2017lectures}. The purposes of identifying these various manifestations of quantumness are manifold. From a theoretical viewpoint, it is crucial to explore the classical-quantum boundary and the quantum origins of our everyday classical world~\cite{zurek2003decoherence}. From a pragmatic perspective, all such forms of quantumness represent resources for some operational tasks and allow us to achieve them with an efficiency that is unreachable by any classical means~\cite{dowling2003quantum}.

In particular, quantum correlations beyond entanglement can be linked to the figure of merit in several operational taks such as local broadcasting~\cite{piani2008no,brandao2015generic}, entanglement distribution~\cite{chuan2012quantum,streltsov2012quantum}, quantum state merging~\cite{cavalcanti2011operational,madhok2011interpreting,streltsov2015concentrating}, quantum state redistribution~\cite{wilde2015multipartite}, quantum state discrimination~\cite{spehner2013geometric,spehner2013geometric2,spehner2014quantum,weedbrook2013discord,farace2014discriminating,roga2015device}, black box quantum parameter estimation~\cite{girolami2014quantum}, quantum data hiding~\cite{piani2014quantumness}, entanglement activation~\cite{piani2011all,gharibian2011characterizing,piani2012quantumness,adesso2014experimental}, device-dependent quantum cryptography~\cite{pirandola2014quantum,pirandola2017fundamental}, quantum work extraction~\cite{horodecki2005local,dillenschneider2009energetics,park2013heat,leggio2015quantum}, quantum refrigeration \cite{correa2013performance,liuzzo2016thermodynamics} and quantum predictive processes \cite{grimsmo2013quantum}.

The quantification of quantum correlations is thus necessary to gauge the quantum enhancement
when performing the aforementioned operational tasks. An intuitive way to measure the quantum correlations present in a state is to quantify the extent to which it violates a property characterising classically correlated states. For example, quantum correlated states cannot be expressed as a statistical mixture of locally, classically distinguishable states~\cite{dakic2010necessary,modi2010unified,paula2013geometric,spehner2013geometric,spehner2013geometric2,roga2016geometric}; are altered by any local measurement~\cite{ollivier2001quantum,henderson2001classical,luo2008using,horodecki2005local}, any non-degenerate local unitary~\cite{giampaolo2013quantifying,roga2014discord} and any local entanglement-breaking channel~\cite{seshadreesan2015fidelity}; always lead to creation of entanglement with an apparatus during a local measurement~\cite{streltsov2011linking,piani2011all,gharibian2011characterizing,piani2012quantumness,adesso2014experimental}; manifest quantum asymmetry with respect to all local non-degenerate observables~\cite{girolami2013characterizing,girolami2014quantum} and coherence with respect to all local bases~\cite{bromley2015frozen,ma2016converting,adesso2016measures}.

The ensuing measures of quantum correlations mostly belong to the following two categories: (i) asymmetric quantifiers, also known as one-sided measures, which vanish only on classical-quantum (resp., quantum-classical) states and thus capture the quantum correlations with respect to subsystem $A$ (resp., $B$) only; (ii) symmetric quantifiers which vanish only on classical-classical states and thus capture the quantum correlations with respect to \textit{either} subsystem $A$ \textit{or} $B$. The latter category of measures have also been improperly referred to as two-sided quantifiers, even though they do not actually capture the quantum correlations with respect to \textit{both} subsystems $A$ \textit{and} $B$.

In this paper we instead introduce an infinite family of quantifiers of quantum correlations beyond entanglement which vanish on both classical-quantum and quantum-classical states and thus properly capture the quantum correlations with respect to both subsystems. More precisely, the `quantum $f-$correlations' are here defined as the maximum metric-adjusted $f-$correlations between pairs of local observables with the same fixed equispaced spectrum and are in one-to-one correspondence with the family of metric-adjusted skew informations \cite{morozova1991markov,petz1996monotone,petz2002covariance,hansen2008metric,gibilisco2007imparato,gibilisco2009imparato,gibilisco2009quantum}. While similar ideas were explored earlier in \cite{davis2000covariance,luo2005quantum} to quantify entanglement, here we show that our quantifiers only reduce to entanglement monotones when restricted to pure states. The latter property is one of the desiderata for general measures of quantum correlations beyond entanglement \cite{adesso2016measures}. Other desiderata, such as monotonicity under sets of operations which cannot create quantum correlations, are also critically assessed. We find in particular that the quantum $f-$correlations, while endowed with strong physical motivations, are not monotone under all such operations in general, although we show in the concluding part of the paper that their definition may be amended to cure this potential drawback.

The paper is organised as follows. In Section \ref{sec:quantumcorrelations} we briefly review the characterisation and quantification of quantum correlations beyond entanglement by adopting a resource-theoretic framework. In Section \ref{sec:maximumquantumcovariance} we define the quantum $f-$correlations and show that they vanish on both classical-quantum and quantum-classical states and are invariant under local unitiaries for any bipartite quantum system. We further prove that they are entanglement monotones when restricted to pure states of qubit-qudit systems. We also analytically evaluate these quantifiers for two-qubit systems and analyse their behaviour under local commutativity preserving channels, showing that they are not monotone in general.
In Section \ref{sec:applications} we provide a physical interpretation for the special quantifier corresponding to the average of the Wigner-Yanase-Dyson skew informations and explore applications to statistical mechanics and many-body systems. We draw our conclusions and outline possible extensions of this work in Section \ref{sec:conclusions}, including a more general definition for a class of quantifiers of two-sided quantum correlations based on  the metric-adjusted $f-$correlations. The latter quantities are proven in Appendix~\ref{sec:appendix} to be monotone under local commutativity preserving channels for two-qubit systems, hence fulfilling \textit{all} the resource-theoretic requirements for quantum correlations beyond entanglement.

\section{Quantifying quantum correlations beyond entanglement}\label{sec:quantumcorrelations}

In this Section we concisely review  the theory of the quantification of quantum correlations beyond entanglement by resorting to a resource-theoretic perspective \cite{coecke2014mathematical,brandao2015reversible}, even though the resource theory of this sort of correlations is still far from being established \cite{horodecki2013quantumness,cianciaruso2015universal}.

From a minimalistic viewpoint, a resource theory relies on the following two ingredients: the sets of free states and free operations, both of which are considered to be freely implementable and are thus such that no resourceful state can be prepared through free operations. A fundamental question that any resource theory must address is how to quantify the resource present in any state. One could naively think that there should be a unique quantifier of a given resource, determining a universal ordering of the resourceful states. However, this should not be the case for the following two reasons. First, the same resource can be exploited for different operational tasks, such that a given resourceful state can be more successful than another one in order to achieve a given operational task, and viceversa when considering another task. Second, it is desirable to assign an operational meaning to any quantifier of a resource, in the sense that it needs to quantify how much the resource possessed by a given state will be useful for achieving a given operational task. An immediate consequence is that, in general, the various quantifiers disagree on the ordering of the resourceful states. Nevertheless, in order to have an operational significance, any {\em bona fide} quantifier of a resource must be compatible with the sets of free states and free operations in the following sense: it must be zero for any free state and monotonically non-increasing under free operations.

Let us start by identifying the set of free states corresponding to quantum correlations beyond entanglement. As we have already mentioned in Section \ref{sec:introduction}, there are at least four settings that we can consider. Within the asymmetric/one-sided setting, when measuring quantum correlations with respect to subsystem $A$ only, the free states are the so-called \textit{classical-quantum} (CQ) \textit{states}, i.e. particular instances of biseparable states than can be written as follows
\begin{equation}\label{CQstate}
\chi_{cq}^{AB} = \sum_i p_i^A \ket{i}\bra{i}^A \otimes \tau_i^B,
\end{equation}
where $\{p_i^A\}$ is a probability distribution,  $\{\ket{i}^A\}$ denotes an orthonormal basis for subsystem $A$, and $\{\tau_i^B\}$ are arbitrary states for subsystem $B$. CQ states represent the embedding of a classical probability distribution $\{p_i^A\}$ relating to only subsystem $A$ into the quantum state space of a bipartite quantum system $AB$.

Analogously, when measuring quantum correlations with respect to subsystem $B$ only, the free states are the so-called \textit{quantum-classical} (QC) \textit{states}, which are of the form
\begin{equation}\label{QCstate}
\chi_{qc}^{AB} = \sum_j p_j^B \tau_j^A \otimes \ket{j}\bra{j}^B,
\end{equation}
where $\{p_j^{B}\}$ is a probability distribution,  $\{\ket{j}^B\}$ denotes an orthonormal basis for subsystem $B$, and $\{\tau_j^A\}$ are arbitrary states for subsystem $A$.

Within the symmetric setting, and when measuring quantum correlations with respect to either subsystem $A$ or $B$, the free states are the so-called \textit{classical-classical} (CC) \textit{states} that can be written in the following form
\begin{equation}\label{CCstate}
\chi_{cc}^{AB} = \sum_{i,j} p_{ij}^{AB} \ket{i}\bra{i}^A \otimes \ket{j}\bra{j}^B,
\end{equation}
where $p_{ij}^{AB}$ is a joint probability distribution, while  $\{\ket{i}^A\}$ and $\{\ket{j}^B\}$ denote orthonormal bases for subsystem $A$ and $B$, respectively. CC states correspond to the embedding of classical bipartite probability distributions $\{p_{ij}^{AB}\}$ into a bipartite quantum state space.

Finally, within the symmetric and properly two-sided setting, wherein one measures quantum correlations with respect to both subsystems $A$ and $B$, the free states are given by the union of the sets of CQ and QC states.

While the free states of the resource theory of general quantum correlations are well identified, the corresponding free operations are still under debate. Having said that, in~\cite{hu2012necessary} it has been shown that all, and only, the \textit{local} operations that leave the set of CQ states invariant are the local commutativity preserving operations (LCPOs) on subsystem $A$, $\Phi^{LCPO}_A \equiv \Lambda_A\otimes \mathbb{I}_B$,  where $\Lambda_{A}$ acts on subsystem $A$ is such a way that $[\Lambda_{A}(\rho^{A}),\Lambda_{A}(\sigma^{A})]=0$ when $[\rho^{A},\sigma^{A}]=0$ for arbitrary marginal states $\rho^{A}$ and $\sigma^{A}$. Analogously, in~\cite{hu2012necessary} it has been also shown that the LCPOs on subsystem $B$, $\Phi^{LCPO}_{B}$, are all and only the local operations leaving the set of QC states invariant, while the LCPOs on both subsystems $A$ and $B$, $\Phi^{LCPO}_{AB}$, are all and only the local operations preserving the set of CC states.  Consequently, due to the fact that free operations cannot create a resourceful state out of a free state, the free operations of the resource theory of quantum correlations beyond entanglement must be within the set of LCPOs, if one imposes a priori the locality of such free operations.

In the case of a qubit, the commutativity preserving operations are constituted by unital and semi-classical channels~\cite{streltsov2011behavior}. Unital channels are defined as those maps that leave the maximally mixed state invariant, whereas semi-classical channels transform the set of all states into a subset of states which are all diagonal in the same basis. More generally, for higher dimensional quantum systems, the commutativity preserving operations are either isotropic or completely decohering channels~\cite{guo2013necessary}.

By considering a resource theory of general quantum correlations corresponding to the largest possible set of local free operations, and taking into account that entanglement is the only kind of quantum correlations that pure states can have, we define any non-negative function  $Q$ on the set of states $\rho$ to be a {\em bona fide} quantifier of two-sided quantum correlations beyond entanglement if it satisfies the following desiderata:
\begin{itemize}
\item(Q1) $Q(\rho)=0$ if $\rho$ is either CQ or QC;
\item(Q2) $Q$ is invariant under local unitaries, i.e. $Q \left((U_A\otimes U_B) \rho (U_A^\dagger \otimes U_B^\dagger) \right)=Q(\rho)$ for any state $\rho$ and any local unitary operation $U_A$ $(U_B)$ acting on subsystem $A$ ($B$);
\item(Q3) $Q (\Phi^{LCPO}_{AB}(\rho)) \leq Q(\rho)$ for any LCPO $\Phi^{LCPO}_{AB}$ on both subsystems $A$ and $B$;
\item(Q4) $Q$ reduces to an entanglement monotone when restricted to pure states.
\end{itemize}
We remark that, while (Q1), (Q2) and (Q4) are well established requirements,  (Q3) may be too strong to impose, as monotonicity under a smaller set of free operations might be sufficient if justified on physical grounds. We will discuss this point further in the following.

For completeness, let us mention that when considering an asymmetric/one-sided measure with respect to subsystem $A$ (resp., $B$), two of the above desiderata have to be slightly modified. Specifically, property (Q1) becomes: $Q(\rho)=0$ if $\rho$ is a CQ (resp., QC) state, while an even stricter monotonicity requirement may replace (Q3), namely being monotonically non-increasing under LCPOs on subsystem $A$ (resp., $B$) and arbitrary local operations on subsystem $B$ (resp., $A$). When considering instead symmetric measures with respect to either subsystem $A$ or $B$, property (Q3) stays the same, while property (Q1) becomes: $Q(\rho)=0$ if $\rho$ is a CC state. On the other hand, properties (Q2) and (Q4) apply equally to all of the aforementioned four settings.

\section{Quantum $f-$correlations}\label{sec:maximumquantumcovariance}

In this Section we define the family of `quantum $f-$correlations' and show that they all satisfy requirements (Q1) and (Q2) for any bipartite quantum system as well as property (Q4) for any qubit-qudit system. We also evaluate these quantifiers in closed form for two-qubit systems and discuss their behaviour under LCPOs, which reveals violations to (Q3), even though these violations can be cured by a suitable reformulation as shown in Section~\ref{sec:extraext}.

\subsection{Metric-adjusted skew informations}

Let us start by introducing the family of metric-adjusted skew informations (MASIs). The Petz classification theorem provides us with a characterisation of the MASIs \cite{morozova1991markov,petz1996monotone,petz2002covariance,hansen2008metric,gibilisco2007imparato,gibilisco2009imparato,gibilisco2009quantum}, by establishing a one-to-one correspondence between them and the Morozova-\v{C}encov (MC) functions
\begin{equation}
 \label{eq:cencovmorozovfunc002}
{c^f}(x,y) = \frac{1}{yf(x/y)}
\end{equation}
parametrized by any function $f(t):\mathbb{R}_+\rightarrow \mathbb{R}_+$ that is
\begin{itemize}
\item{($i$) operator monotone (or \textit{standard}), i.e. for any positive semi-definite operators $A$ and $B$ such that $A \leq B$, then $f(A) \leq f(B)$;}
\item{($ii$) symmetric (or \textit{self-inversive}), i.e. $f(t) = tf(1/t)$;}
\item ($iii$) normalised, i.e. $f(1) = 1$.
\end{itemize}
The set of all normalised symmetric operator monotone functions $f$ on the interval $(0, + \infty)$ is usually denoted by ${\cal F}_{op}$.
It follows that any MC function is symmetric in its arguments, i.e. ${c^f}(x,y) = {c^f}(y,x)$, and homogeneous of degree $-1$, i.e. ${c^f}(\alpha x,\alpha y) = {\alpha^{-1}}{c^f}(x,y)$.

In this formalism, the MASI of a quantum state $\rho>0$ with respect to an observable $O$, corresponding to the MC function $c^f$, can be defined as follows \cite{hansen2008metric}:
\be\label{eq:def_MASI}
 	I^f(\rho, O) = \frac{f(0)}{2} \sum_{ij} c^f(p_i,p_j) (p_i - p_j)^2 \langle i \vert O \vert j \rangle  \langle j \vert O \vert i \rangle,
 \ee
where $\rho = {\sum_i}{p_i}|{i}\rangle\langle{i}|$ is the spectral decomposition of $\rho$ and we have assumed $f$ to be {\em regular}, i.e. $\lim_{t\rightarrow 0^+}f(t) \equiv f(0)>0$. Notable examples of MASIs are the Bures-Uhlmann information \cite{uhlmann1976transition}, corresponding to the maximal function $f^{BU}(t)=(1+t)/2$, and the Wigner-Yanase-Dyson skew informations \cite{wigner1963information}, corresponding to the functions
\begin{equation}
\label{eq:WYD}
f^{WYD}_\alpha(t)=\frac{\alpha (1-\alpha)(1-t)^2}{(1-t^\alpha)(1-t^{1-\alpha})},
\end{equation}
for any $0<\alpha<1$.

Each MASI $I^f(\rho, O)$ can be interpreted as a genuinely quantum contribution to the uncertainty of the observable $O$ in the state $\rho$ \cite{petz2002covariance, luo2005quantum, luo2006quantum, gibilisco2007imparato, gibilisco2009imparato, gibilisco2009quantum, li2011averaged}. Two important properties of $I^f(\rho, O)$, justifying this intuition, are that: (a) $I^f(\rho, O) = 0$ iff $[\rho, O] = 0$; and (b) $I^f(\rho, O) \le {\rm Var}_{\rho}(O) = {\rm Tr}(\rho O^2) -  [{\rm Tr}(\rho O)]^2$ where the equality holds for pure states. Hence, a nonzero MASI indicates that the state $\rho$ contains coherences among different eigenstates of $O$ [property (a)]. For a pure state, any source of uncertainty has a quantum origin, and all MASIs coincide with the ordinary (Robertson-Schr\"odinger) variance of $O$ in the state.
%
%

The MASI $I^f(\rho, O)$ may also be interpreted as asymmetry of the state $\rho$ with respect to the observable $O$ \cite{marvian2014extending, girolami2014observable, zhang2016determining}.
In a bipartite system $\rho_{AB}$, the minimum of $I^f(\rho_{AB}, O_A\otimes\mathbb{I}_B)$ over local non-degenerate observables $O_A$ (with fixed spectrum) can be seen as a measure of asymmetric/one-sided quantum correlations of the state $\rho_{AB}$ with respect to  subsystem $A$, as investigated for special instances in~\cite{girolami2013characterizing,girolami2014quantum}. Applications of different MASIs to determining quantum speed limits for closed and open quantum system dynamics have been explored in \cite{pires2016generalized,marvian2016quantum} and references therein.

\subsection{Maximising metric-adjusted $f-$correlations over pairs of local observables}

We now recall the notion of {\em metric-adjusted $f-$correlations} between observables $O_A$ and $O_B$ in the quantum state $\rho = {\sum_i}{p_i}|{i}\rangle\langle{i}|$, defined by \cite{hansen2008metric,gibilisco2007imparato}
\be \label{eq:defqCOV}
	{\Upsilon}^f(\rho, O_A, O_B) =  \frac{f(0)}{2} \sum_{ij} c^f(p_i,p_j)(p_i - p_j)^2 \langle i \vert O_A \vert j \rangle  \langle j \vert O_B \vert i \rangle,
\ee
Equivalently, one can write \cite{gibilisco2011isola,audenaert2008inequalities}
\be \label{eq:defqCOV2}
	{\Upsilon}^f(\rho, O_A, O_B) =  {\rm Cov}_\rho(O_A,O_B) - {\rm Cov}^{\tilde{f}}_\rho(O_A,O_B),
\ee
where
\be\label{eq:fcovariance}
{\rm Cov}^{f}_\rho(O_A,O_B) \!=\! {\rm Tr}\left\{m^f\left[\rho \big(O_A\!-\!{\rm Tr}(\rho O_A)\big), \big(O_A\!-\!{\rm Tr}(\rho O_A)\big)\rho\right] \big(O_B\!-\!{\rm Tr}(\rho O_B)\big)\right\}
\ee
stands for the Petz $f-$covariance \cite{petz2002covariance} associated with the Kubo-Ando operator mean $m^f[A,B]=A^{\frac12}f(A^{-\frac12}BA^{\frac12})A^{\frac12}$ \cite{kubo1980means}, reducing to the ordinary (Robertson-Schr\"odinger) covariance
\be\label{eq:covariance}
{\rm Cov}_\rho(O_A,O_B) =\frac12 {\rm Tr}\left[\rho (O_A O_B + O_B O_A)\right] - {\rm Tr}(\rho O_A) {\rm Tr}(\rho O_B)
\ee
for $f(t) \equiv f^{BU}(t)=(1+t)/2$ (in which case $m^f$ denotes the arithmetic mean), and  \cite{gibilisco2009correspondence,gibilisco2011isola}
\be\label{eq:ftilde}
\tilde{f}(t) = \frac12\left[(t+1)-(t-1)^2\frac{f(0)}{f(t)} \right],
\ee
for any regular $f \in {\cal F}_{op}$.
It follows from Eq.~(\ref{eq:defqCOV2}) or, alternatively, from Eq.~(\ref{eq:defqCOV}) due to the symmetry of the MC functions $c^f(p_i,p_j)$,
 that
\be \label{eq:nonadditivitydefinition}
	I^f(\rho, O_A + O_B) = I^f(\rho, O_A) + I^f(\rho, O_B) + 2 {\Upsilon}^f(\rho, O_A, O_B).
\ee
In other words, the metric-adjusted $f-$correlations can be seen as measures of non-additivity of the corresponding MASIs.

We are now ready to define the {\em quantum $f-$correlations} of a state $\rho$ as
\begin{equation}\label{eq:LQCdefinition}
{\rm Q}^{f}(\rho) =  \max_{O_A,O_B} {\Upsilon}^f(\rho,O_A\otimes\mathbb{I}_B,\mathbb{I}_A\otimes O_B),
\end{equation}
where the maximisation is over all local observables $O_A$ and $O_B$ whose eigenvalues are equispaced with spacing $d/(d-1)$ and are given by $\{-d/2,-d/2+d/(d-1),\cdots,d/2-d/(d-1), d/2\}$, with $d=\min\{d_A,d_B\}$. If the dimensions of the two subsystems are different, say $d_B>d_A$, the remaining eigenvalues of $O_B$ are set to zero (and vice-versa if $d_A>d_B$).

\subsection{The quantum $f-$correlations satisfy Q1 and Q2}\label{sec_qcovar_vanishes_on_QC_CQ_states}

We now show that the quantity ${\rm Q}^{f}$ defined in Eq.~(\ref{eq:LQCdefinition}) vanishes on both CQ and QC states and thus satisfies requirement (Q1) for any function $f \in {\cal F}_{op}$.  This is due to the fact that the metric-adjusted $f-$correlations actually vanish on both CQ and QC states for any pair of local observables $O_A$ and $O_B$. Indeed, consider a CQ state as in Eq.~(\ref{CQstate}). This can also be written as follows:
\be \label{eq:CQstateDecomposed}
		\chi_{cq}^{AB} = \sum_{i, j} p_{i, j} \vert i^A \rangle  \langle i^A \vert \otimes \vert \psi_{i,j}^B \rangle   \langle \psi_{i,j}^B \vert,
	\ee
where we have used the spectral decomposition of the states $\tau_i^B$, i.e. $\tau_i^B = \sum_j q_{j | i} \vert \psi_{i,j}^B \rangle \langle \psi_{i,j}^B \vert$, and introduced the probabilities $p_{i, j} = p^A_i q_{j | i}$.

By using Eq.~(\ref{eq:CQstateDecomposed}) we can see that (up to the factor $f(0)/2$)
	\bearr
		&&\!\!\!\!\!\!\!\!{\Upsilon}^f(\chi_{cq}^{AB}, O_A\otimes \mathbb{I}_B, \mathbb{I}_A \otimes O_B) \\
		&&\propto \sum_{i, j, k, l} g^f(p_{i, j},p_{k, l})\langle i^A \vert \langle \psi_{i,j}^B \vert O_A\otimes \mathbb{I}_B \vert k^A \rangle \vert \psi_{k,l}^B \rangle  \langle k^A \vert \langle \psi_{k, l}^B \vert \mathbb{I}_A \otimes O_B \vert i^A \rangle \vert \psi_{i, j}^B \rangle \nonumber \\
		&&= \sum_{i, j, k, l} g^f(p_{i, j},p_{k, l})\langle i^A \vert O_A \vert k^A \rangle \langle \psi_{i,j}^B \vert \psi_{k,l}^B \rangle  \langle k^A \vert i^A \rangle \langle \psi_{k, l}^B \vert O_B \vert   \psi_{i, j}^B \rangle \nonumber \\
		&&= \sum_{i, j, l} g^f(p_{i, j},p_{i, l})\langle i^A \vert O_A \vert i^A \rangle \langle \psi_{i,j}^B \vert \psi_{i,l}^B \rangle   \langle \psi_{i, l}^B \vert O_B \vert   \psi_{i, j}^B \rangle \nonumber \\
		&&= \sum_{i, j} g^f(p_{i, j},p_{i, j})\langle i^A \vert O_A \vert i^A \rangle \langle \psi_{i, j}^B \vert O_B \vert   \psi_{i, j}^B \rangle  \nonumber \\
		&&= 0, \nonumber
	\eearr
being $\langle k^A \vert i^A \rangle = \delta_{i,k}$, $\langle \psi_{i,j}^B \vert \psi_{i,l}^B \rangle = \delta_{j,l}$ for any $i$ and $g^f(p_{i, j},p_{i, j}) \equiv c^f(p_{i, j},p_{i, j})(p_{i, j} - p_{i, j})^2 = 0$. An analogous reasoning applies when considering QC states, thus concluding our proof.

It is also clear that ${\rm Q}^{f}$ is by construction invariant under local unitaries, as the latter cannot vary the spectrum of the local observables involved in the optimisation in Eq.~(\ref{eq:LQCdefinition}), so that ${\rm Q}^{f}$ satisfies requirement (Q2) for any bipartite system.

\subsection{Quantum $f-$correlations as entanglement monotones for pure qubit-qudit states}

Specialising our discussion to qubit-qudit systems, we now show that ${\rm Q}^{f}$ is an entanglement monotone \cite{vedral1997quantifying,vidal2000entanglement} when restricted to pure states, and thus satisfies requirement (Q4) for this special class of bipartite systems. For every MC function $c^f$, the quantity ${\rm Q}^{f}$ reduces to the maximum ordinary (Robertson-Schr\"odinger) covariance of local observables when calculated for pure states, i.e.
\begin{equation}\label{eq:entanglementcovariance}
{\rm Q}^{f}(|\psi\rangle) = E(|\psi\rangle) \equiv \max_{O_A,O_B} (\langle\psi| O_A\otimes O_B |\psi\rangle - \langle\psi| O_A\otimes \mathbb{I}_B|\psi\rangle\langle\psi| \mathbb{I}_A\otimes O_B|\psi\rangle),
\end{equation}
where the maximisation is over all local observables $O_A$ and $O_B$ with equispaced eigenvalues. We thus want to prove that this is a pure state entanglement monotone.

It is known that if $E(|\psi\rangle)$ can be written as a Schur-concave function of the Schmidt coefficients $\{\lambda_i\}$ of $|\psi\rangle$, then $E(|\psi\rangle)$ is a pure state entanglement monotone \cite{nielsen1999conditions,nielsen2001majorization}. Let us recall that the Schmidt decomposition \cite{nielsen2010quantum} of a bipartite pure state $|\psi\rangle$ is given by
\begin{equation}
|\psi\rangle = \sum_{i=1}^d \lambda_i |e_i^A\rangle\otimes|f_i^B\rangle,
\end{equation}
where $\{|e_i^A\rangle\}$ and $\{|f_i^B\rangle\}$ are orthonormal states of subsystems $A$ and $B$, and the Schmidt coefficients $\lambda_i$ satisfy $\lambda_i \geq 0$ and $\sum_{i=1}^d \lambda_i^2 = 1$.

By substituting the Schmidt decomposition of $|\psi\rangle$ into Eq.~(\ref{eq:entanglementcovariance}) we get
\begin{equation}\label{eq:entanglementcovarianceintermsofSchmidt}
E(|\psi\rangle) = \sum_{i,j=1}^d a_{ij}b_{ij} \lambda_i\lambda_j - \sum_{i,j=1}^d a_{ii}b_{jj} \lambda_i^2\lambda_j^2,
\end{equation}
where $a_{ij}=\langle e_i^A|O_A^*| e_j^A\rangle$, $b_{ij}=\langle f_i^B|O_B^*| f_j^B\rangle$, while $O_A^*$ and $O_B^*$ are the local observables achieving the maximum in Eq.~(\ref{eq:entanglementcovariance}). Moreover, by using the fact that $\sum_{i=1}^d \lambda_i^2=1$, we have that
\begin{equation}\label{eq:entanglementcovarianceintermsofSchmidtsimplified}
E(|\psi\rangle) = \sum_{j>i=1}^d(a_{ii} - a_{jj})(b_{ii} - b_{jj}) \lambda_i^2\lambda_j^2 + \sum_{j>i=1}^d(a_{ij}b_{ij} + a_{ji}b_{ji}) \lambda_i\lambda_j.
\end{equation}

Now we just need to prove that the function expressed in Eq.~(\ref{eq:entanglementcovarianceintermsofSchmidtsimplified}) is Schur-concave. The Schur-Ostrowski criterion \cite{ando1989majorization} says that a symmetric function $f(\lambda_1,\lambda_2,\cdots,\lambda_d)$ is Schur-concave if, and only if,
\begin{equation}
(\lambda_i - \lambda_j) \left( \frac{\partial f}{\partial \lambda_i} - \frac{\partial f}{\partial \lambda_j} \right) \leq 0
\end{equation}
for any $j > i \in \{1,2,\cdots, d\}$. However, before applying this criterion to the function in Eq.~(\ref{eq:entanglementcovarianceintermsofSchmidtsimplified}), we need to find the optimal local observables $O_A^*$ and $O_B^*$ and thus the explicit form of the coefficients $a_{ij}$ and $b_{ij}$.

We may now exploit some convenient simplifications occurring in the case $d=2$, i.e., for any qubit-qudit bipartite system. In this case we have that
\begin{equation}\label{eq:entanglementcovarianceintermsofSchmidtd2}
E(|\psi\rangle) = (a_{11} - a_{22})(b_{11} - b_{22}) \lambda_1^2\lambda_2^2 + (a_{12}b_{12} + a_{21}b_{21}) \lambda_1\lambda_2,
\end{equation}
which is a symmetric function of $\lambda_1$ and $\lambda_2$, regardless of the form of the local optimal observables $O_A^*$ and $O_B^*$. Having $O_A^*$ and $O_B^*$ the same spectrum $\{-1,1\}$ by construction, we can easily see that the optimal local observables must be of the form
\begin{equation}
O_A^* =
\left(\begin{array}{ccccc}
a & e^{i \varphi} \sqrt{1-a^2} \\
e^{-i \varphi} \sqrt{1-a^2} & -a \\
\end{array}\right),
\end{equation}
\begin{equation}
O_B^* =
\left(\begin{array}{ccccc}
b & e^{i \phi} \sqrt{1-b^2} & 0 & \cdots & 0\\
e^{-i \phi} \sqrt{1-b^2} & -b & 0 & \cdots & 0\\
0 & 0 & 0 & \cdots & 0\\
\vdots & \vdots &  \vdots &\ddots & \vdots\\
0 & 0 & 0 & \cdots & 0\\
\end{array}\right),
\end{equation}
for some $-1 \leq a\leq 1$ and $-1 \leq b\leq 1$, so that Eq.~(\ref{eq:entanglementcovarianceintermsofSchmidtd2}) becomes
\begin{equation}\label{eq:entanglementcovarianceintermsofSchmidtd2simplified}
E(|\psi\rangle) = 4 a b \lambda_1^2\lambda_2^2 + 2 \sqrt{(1-a^2)(1-b^2)} \cos(\varphi+\phi) \lambda_1\lambda_2,
\end{equation}
whose maximum is given by $\varphi=\phi=a=b=0$, i.e.,
\begin{equation}\label{eq:entanglementcovarianceintermsofSchmidtd2final}
E(|\psi\rangle) = 2 \lambda_1\lambda_2.
\end{equation}

By applying the Schur-Ostrowski criterion to the function $E$ in Eq.~(\ref{eq:entanglementcovarianceintermsofSchmidtd2final}), we find that
\begin{equation}
(\lambda_1 - \lambda_2) \left( \frac{\partial E}{\partial \lambda_1} - \frac{\partial E}{\partial \lambda_2} \right) = - 2 (\lambda_1 - \lambda_2)^2 \leq 0,
\end{equation}
so that $E(|\psi\rangle)$ is a Schur-concave function of the Schmidt coefficients of $|\psi\rangle$ and thus is a pure state entanglement monotone for any qubit-qudit system. In particular, for a qubit-qubit system the above function reduces to the well known concurrence \cite{hill1997entanglement}.

\subsection{Analytical expression of the two-qubit quantum $f-$correlations}

We now analytically evaluate ${\rm Q}^{f}(\rho)$ for any MC function $c^f$ when restricting to two-qubit states. We start by noting that in the two-qubit case any local observable whose spectrum is given by $\{-1,1\}$ can be written as $O=\bm{n}\cdot \bm{\sigma}$, with $\bm{n}=\{n_1,n_2,n_3\}$ being a real unit vector, $\bm{n} \cdot \bm{n} = 1$, and $\bm{\sigma}=\{\sigma_1,\sigma_2,\sigma_3\}$ is the vector of Pauli matrices. Therefore, Eq.~(\ref{eq:LQCdefinition}) becomes
\begin{equation}
{\rm Q}^{f}(\rho^{AB}) = \max_{\bm{n}_A,\bm{n}_B} \bm{n}^T_A M^f \bm{n}_B,
\end{equation}
where the maximum is over all real unit vectors $\bm{n}_A$ and $\bm{n}_B$, while $M^f$ is the $3\times3$ matrix with elements
\begin{equation}\label{eq:WYcorrelationmatrix}
M_{ij}^f = {\Upsilon}^f(\rho,\sigma_i^A\otimes\mathbb{I}_B,\mathbb{I}_A\otimes \sigma_j^B),
\end{equation}
so that we can formally write the result of the maximisation as
\begin{equation}
{\rm Q}^{f}(\rho^{AB}) = s_{\max}(M^f),
\end{equation}
where $s_{\max}(M^f)$ is the maximum singular value of the matrix $M^f$.

\subsection{Behaviour of the quantum $f-$correlations under local commutativity preserving channels}

\begin{figure}[t!]
    \centering
    \includegraphics[width=8cm]{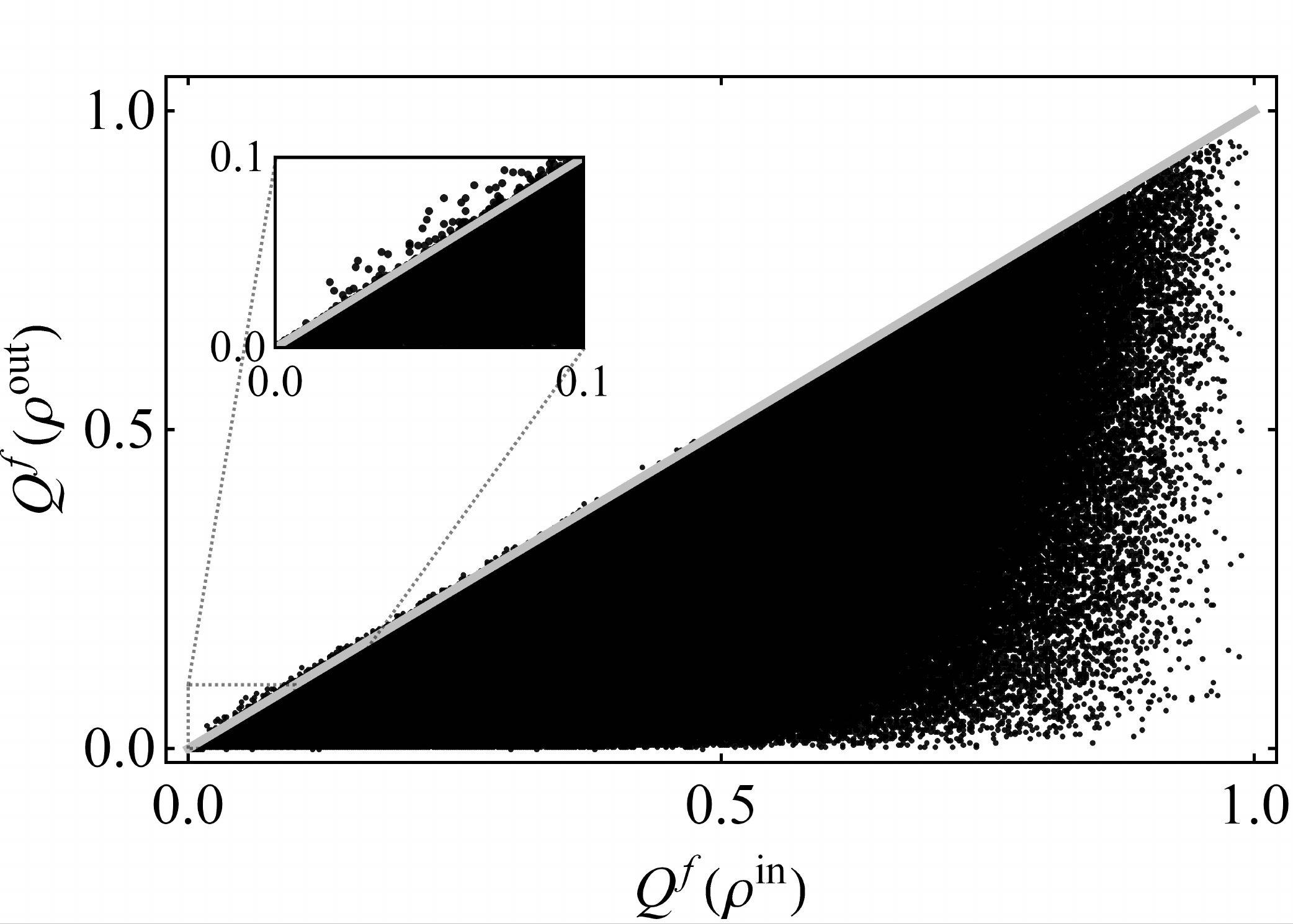}
    \caption{Comparison between the quantum $f-$correlation ${\rm Q}^{f}(\rho^{\text{in}})$ of $10^6$ random two-qubit states (horizontal axis), and the quantum $f-$correlation ${\rm Q}^{f}(\rho^{\text{out}})$ of the corresponding states after random local unital channels (vertical axis), for  $f(t)=(1+\sqrt{t})^2/4$ associated with the Wigner-Yanase skew information. The presence of points above the solid gray line ${\rm Q}^{f}(\rho^{\text{out}})={\rm Q}^{f}(\rho^{\text{in}})$, better highlighted in the zoomed-in inset, shows that the quantum $f-$correlations are in general not monotonically non-increasing under local unital channels for two qubits, as discussed further in the main text.}
    \label{fig:viola}
\end{figure}

Here we investigate whether the quantum $f-$correlations defined in Eq.~(\ref{eq:LQCdefinition}) are monotonically nonincreasing under LCPOs, as would be demanded by the resource-theoretic desideratum (Q3). The answer is trivially affirmative in the case of local semi-classical channels, which map any state into one with vanishing ${\rm Q}^{f}$. To investigate the non-trivial cases, we carry out a numerical exploration for two-qubit states subject to local unital channels. Fig.~\ref{fig:viola} compares the input ${\rm Q}^{f}(\rho^{\text{in}})$ with the output ${\rm Q}^{f}(\rho^{\text{out}})$ for $10^6$ randomly generated two-qubit states $\rho^{\text{in}}$, where $\rho^{\text{out}}= p (U_A \otimes \mathbb{I}_B) \rho^{\text{in}} (U_A^\dagger \otimes \mathbb{I}_B) + (1-p) (V_A \otimes \mathbb{I}_B) \rho^{\text{in}} (V_A^\dagger \otimes \mathbb{I}_B)$, with random local unitaries $U_A, V_A$, and random probability $p \in [0,1]$. The analysis reported in Fig.~\ref{fig:viola} has been done in particular using the Wigner-Yanase skew information \cite{wigner1963information}, specified by $f^{WY} \equiv f^{WYD}_{1/2}$ [see Eq.~(\ref{eq:WYD})], although qualitatively similar results are obtained for other choices of $f$. As clear from the plot, while monotonicity under local unital channels appears to hold in most cases, narrow violations can still be identified in about $0.1\%$ of the cases in our study. This shows that the quantifers defined by Eq.~(\ref{eq:LQCdefinition}) can  increase under some LCPOs, thus generally failing to fulfil (Q3).

On one hand, this may suggest that the quantum $f-$correlations are not entirely satisfactory measures of general quantum correlations from a resource theory perspective, while still providing an approximately reliable quantitative estimate. On the other hand, this may indicate that a more narrow and possibly physically relevant subset of LCPOs may play a preferred role in identifying the free operations for the resource theory of general quantum correlations, and the quantum $f-$correlations could still be monotone under such a restricted set.

In fact, the latter scenario resembles what happens in the resource theories of entanglement and coherence, wherein the chosen free operations do not cover the whole maximal set of operations leaving the set of free states invariant. For example, in the entanglement case, the free operations are the local operations and classical communication, which are only a restricted subset of the separability preserving operations \cite{chitambar2014everything}. In the coherence case, there are in fact many different definitions of free operations that are proper subsets of the maximal set of incoherence preserving operations \cite{aaberg2006quantifying}, such as the incoherent operations \cite{baumgratz2014quantifying}, the strictly incoherent operations \cite{yadin2016quantum}, the translationally invariant operations \cite{marvian2016quantum}, and several others \cite{streltsov2016quantum}, with no consensus yet reached on the most representative set.

In the case of general quantum correlations, as already mentioned in Section~\ref{sec:quantumcorrelations}, the quest for the physical justification to identify the right set of free operations is still open \cite{adesso2016measures}. Based on our numerical analysis for two-qubit systems, there was no subset of local unital channels which clearly emerged as the one under which monotonicity could hold in general. Hence, a way to save (Q3) for the quantum $f-$correlations could be to impose only monotonicity under local semi-classical channels, which might be nonetheless too weak a constraint.

In Section \ref{sec:applications}, we discuss one possible physical setting that bolsters, from a different perspective, the interpretation of the quantum $f-$correlations as indicators of quantum correlations, leaving aside the critical resource-theoretic characterisation of the ensuing set of free operations. We return to the latter issue in the concluding Section \ref{sec:conclusions}, where an amended definition to cure the drawbacks of Eq.~(\ref{eq:LQCdefinition}) is proposed and validated, leading in particular to generalised quantifiers for which monotonicity under local unital channels does hold for all two-qubit states.

\section{Physical interpretation and applications}\label{sec:applications}

In this Section we provide a physical interpretation for the quantum $f-$correlation corresponding to the average of the Wigner-Yanase-Dyson skew informations, which is itself a member of the family of MASIs, as shown in the following.
As we mentioned above, each MASI $I^f(\rho, O)$ defined in Eq.~(\ref{eq:def_MASI}) can be used to quantify the coherent spread of the state $\rho$ across the eigenstates of an observable $O$ --- or the quantum portion of the total uncertainty ${\rm Var}_{\rho}(O)$.
The metric-adjusted $f-$correlations defined in Eq.~(\ref{eq:nonadditivitydefinition}), which stem from the non-additivity of these MASIs, thus have the transparent meaning of \emph{quantum contributions to the covariance of different observables}. Among the MASIs, one of them takes a special meaning for thermal equilibrium states
\begin{equation}
	\rho = \frac{1}{Z} e^{-H / T} ~,
\end{equation}
where $T$ is the temperature (in natural units) and $H$ the Hamiltonian of the $AB$ system, while the partition function $Z = {\rm Tr} (e^{-H/T})$ ensures that the density matrix is normalised, ${\rm Tr}(\rho)=1$.  Indeed, let us consider the quantity (referred to as `quantum variance' in \cite{frerot2016quantum})
\begin{equation}\label{eq:MASIqv}
	I^{\bar f}(\rho, O) \equiv  \int_{0}^1 d\alpha ~ I^\alpha (\rho, O) ~,
\end{equation}
where $I^\alpha \equiv I^{f_\alpha^{WYD}}$ is the Wigner-Dyson-Yanase skew information \cite{hansen2008metric} with $f_\alpha^{WYD}$ defined in Eq.~(\ref{eq:WYD}). Then, using the methods of \cite{hansen2008metric}, it is straightforward to show that $I^{\bar f}(\rho, O)$ is a MASI in the form of Eq.~(\ref{eq:def_MASI}), defined by the corresponding operator monotone function $\bar{f} \in {\cal F}_{op}$, which takes the expression
\begin{equation}
\bar f(t)=\frac{(1-t)^2}{12 \left(\frac{t+1}{2}-\frac{t-1}{\log (t)}\right)}\,.
\end{equation}
It turns out that, for this instance, the associated metric-adjusted $f-$correlation can be defined, independently of Eq.~(\ref{eq:defqCOV2}), as \cite{malpetti2016quantum}
\be
	{\Upsilon}^{\bar f}(\rho, O_A, O_B) \equiv
	{\rm Cov}(O_A,O_B) - T \left. \frac{\partial \langle O_A \rangle}{\partial h_B} \right|_{h_B = 0} ~.
	\label{eq:defquantumcovariancethermo}
\ee
Here, ${\rm Cov}(O_A,O_B)$ is the ordinary covariance defined in Eq.~(\ref{eq:covariance}), and $\frac{\partial \langle O_A \rangle}{\partial h_B}$ is the static susceptibility of $\langle O_A \rangle$ with respect to the application of a field $h_B$ which couples to $O_B$ in the Hamiltonian, $H(h_B) =  H - h_B O_B$. The equality
\begin{equation}
	{\rm Cov}(O_A,O_B) = T \left. \frac{\partial \langle O_A \rangle}{\partial h_B} \right|_{h_B = 0} ~,
\end{equation}
is a thermodynamic identity (a ``fluctuation-dissipation theorem'' \cite{kubo1966fluctuation}) for classical systems at thermal equilibrium. Therefore, the genuinely quantum contribution to the covariance as defined in Eq.~(\ref{eq:defquantumcovariancethermo}) quantifies those correlations between a pair of local observables $O_A$ and $O_B$ which cannot be accounted for by classical statistical mechanics. As we have proved in Section \ref{sec_qcovar_vanishes_on_QC_CQ_states}, the discrepancy between classical and quantum statistical mechanics  can be traced back, within the framework of quantum information theory, to the state $\rho$ not being CQ or QC.

Defining the nonclassical contribution to the covariance in a thermal state via Eq.~(\ref{eq:defquantumcovariancethermo}) is experimentally and computationally appealing, because one does not rely upon the tomographic reconstruction of the state, \textit{a priori} needed in view of the general definition of Eq.~(\ref{eq:defqCOV}), and which is prohibitive for large systems. Being defined in terms of measurable quantities (namely usual correlation and response functions), Eq.~(\ref{eq:defquantumcovariancethermo})  provides a  convenient tool to access two-sided quantum correlations  in quantum systems at thermal equilibrium. Moreover, Eq.~(\ref{eq:defquantumcovariancethermo}) is accessible also in the case of large-scale numerical calculations: in \cite{malpetti2016quantum}, the spatial structure of the quantity in Eq.~(\ref{eq:defquantumcovariancethermo}) has been investigated for many-body systems of thousands of qubits using quantum Monte Carlo methods. It should also be accessible to state-of-the-art cold-atom experiments as proposed in \cite{frerot2016quantum}.

\section{Discussion and conclusions}\label{sec:conclusions}

We have defined an infinite family of quantitative indicators of two-sided quantum correlations beyond entanglement, which vanish on both classical-quantum and quantum-classical states and thus properly capture quantumness with respect to both subsystems. These quantifiers, named `quantum $f-$correlations', are in one-to-one correspondence with the metric-adjusted skew informations \cite{morozova1991markov,petz1996monotone,petz2002covariance,hansen2008metric,gibilisco2007imparato,gibilisco2009imparato,gibilisco2009quantum}. We have shown that the quantum $f-$correlations are entanglement monotones for pure states of qubit-qudit systems, having also provided closed-form expressions for these quantifiers for two-qubit systems. We further analysed their behaviour under local commutativity preserving operations. Focusing on systems at thermal equilibrium, a situation especially relevant to many-body systems, we have physically interpreted the quantifier corresponding to the average of the Wigner-Yanase-Dyson skew informations by resorting to a quantum statistical mechanics perspective \cite{frerot2016quantum,malpetti2016quantum}.

The still unsolved characterisation of the subset of local commutativity preserving operations under which the quantum $f-$correlations are monotonically nonincreasing deserves special attention in light of the quest for the identification of physically relevant free operations within a resource theory of general quantum correlations \cite{adesso2016measures}. Further investigation towards a deeper understanding of the quantum $f-$correlations for higher dimensional and multipartite quantum systems is also worthwhile. In particular, it could be interesting to explore the possible operational role played by these quantifiers in multiparameter quantum estimation \cite{braunstein1994statistical,giovannetti2006quantum,ragy2016compatibility}.

\subsection{Extending the optimisation in the definition of the quantum $f-$correlations}\label{sec:extraext}

Finally, we note that the notion of two-sided quantum correlations we have introduced depends nontrivially on what portion of a multipartite system is assumed to be accessible. Indeed, if $O_A$ and $O_B$ in Eq.~(\ref{eq:defqCOV}) are local observables acting on two different subsystems $A$ and $B$ of a larger system $ABC$, the quantum covariance between $O_A$ and $O_B$ will in general take a different value if calculated on the full tripartite state $\rho^{ABC}$, as compared to the original state $\rho^{AB}={\rm Tr}_C[\rho^{ABC}]$. Furthermore, we have verified numerically that in general the two quantities do not satisfy a particular ordering. This issue appears to be at the root of the violation of the monotonicity (Q3) for the quantum $f-$correlations under local commutativity preserving operations. In the interest of removing such an ambiguity, we conjecture that a general and {\em bona fide} quantifier of two-sided quantum correlations, solely dependent on the state $\rho^{AB}$, may be defined as follows:
\be \label{eq:defqCOVextended}
\widetilde{{\rm Q}}^{f}(\rho^{AB})\equiv\!\!\!\!\mathop{\inf_{\rho^{ABC} \text{ s.t.}}}_{{\rm Tr}_C[\rho^{ABC}]=\rho^{AB}}\!\!\left[  \max_{O_A,O_B} {\Upsilon}^f(\rho^{ABC},O_A\otimes\mathbb{I}_B\otimes \mathbb{I}_C,\mathbb{I}_A\otimes O_B\otimes \mathbb{I}_C)\right]\!.
\ee
The optimisation problem above, performed over all the possible extensions of the state $\rho^{AB}$ into a larger Hilbert space, appears to be a rather daunting task.
Yet, trading computability for reliability \cite{tufarelli2013geometric}, it is interesting to assess whether the quantity in Eq.~(\ref{eq:defqCOVextended}) may serve as a meaningful tool to provide further insight on the operational interpretation and mathematical characterisation of two-sided quantum correlations beyond entanglement, in particular respecting all desiderata arising from a resource-theoretic approach while maintaining a clear physical motivation.

Here we provide a first affirmative answer.
In particular, we prove in Appendix~\ref{sec:appendix} that $\widetilde{{\rm Q}}^{f}(\rho^{AB})$ is in fact monotonically nonincreasing under all local unital channels for any two-qubit state $\rho^{AB}$, hence fulfilling requirement (Q3) in this prominent instance. A more general investigation into the monotonicity properties of $\widetilde{{\rm Q}}^{f}(\rho^{AB})$ under local commutativity preserving channels (or relevant subsets thereof) for states $\rho^{AB}$ of arbitrary dimension will be the subject of future work.

\section*{Acknowledgments}
We thank Thomas Bromley and Tommaso Roscilde for stimulating discussions, as well as Paolo Gibilisco and an anonymous referee for very fruitful comments on a previous version of this manuscript. We acknowledge financial support from the European Research Council (Grant No.~637352 GQCOP), the Foundational Questions Institute (Grant No.~FQXi-RFP-1601), and the Agence Nationale de la Recherche (``ArtiQ" project). T.T. acknowledges financial support from the University of Nottingham via a Nottingham Research Fellowship.
\appendix
\section{Monotonicity of Eq.~(\ref{eq:defqCOVextended}) under local unital channels\label{sec:appendix}}
We will here prove that, if $\Lambda_A$ is a unital channel on qubit $A$, then the following inequality holds:
\begin{equation}\label{monodavero}
\widetilde{{\rm Q}}^{f}(\Lambda_A(\rho^{AB}))\leq \widetilde{{\rm Q}}^{f}(\rho^{AB}).
\end{equation}
In order to prevent the notation from becoming too cumbersome, in this Appendix we shall leave identity operators implicit wherever convenient: for example in the equation above we defined  $\Lambda_A(\rho^{AB})\equiv\Lambda_A\otimes\mathbb{I}_B(\rho^{AB})$.

To begin our proof, let us assume that $\rho^{ABC}$ is the optimal dilation of $\rho^{AB}$ for the sake of Eq.~(\ref{eq:defqCOVextended}), that is, $\widetilde{{\rm Q}}^{f}(\rho^{AB})={{\rm Q}}^{f}_{AB}(\rho^{ABC})$, where the subscript ${AB}$ indicates what subsystems are involved in the calculation of the relevant quantum $f-$correlations. Consider now any dilation $\tau^{ABCD}$ of $\Lambda_{A}(\rho^{ABC})$ into a larger space, including a further ancillary system $D$. We note that $\tau^{ABCD}$ is automatically also a dilation of $\Lambda_{A}(\rho^{AB})$. Hence, the following inequality holds by definition:
\begin{equation}\label{4mono}
\widetilde{{\rm Q}}^{f}(\Lambda_A(\rho^{AB}))\leq {{\rm Q}}^{f}_{AB}(\tau^{ABCD}),
\end{equation}
Eq.~(\ref{monodavero})  can then be proven by showing that ${{\rm Q}}^{f}_{AB}(\tau^{ABCD})\le {{\rm Q}}^{f}_{AB}(\rho^{ABC})$ for a particular choice of $\tau^{ABCD}$.

To proceed, we use the fact that any unital qubit operation can be equivalently written as a random unitary channel \cite{nielsen2010quantum}, i.e.
\begin{equation}
\Lambda_{A}(\bullet) = \sum_{k} q_{k}\, U_{A}^{(k)} \;\bullet \;( U_{A}^{(k)}) ^{\dagger},
\end{equation}
for an appropriate collection of unitaries $\{U_{A}^{(k)}\}$ (acting on subsystem $A$) and probabilities $\{q_{k}\}$. A suitable dilation of $\Lambda_{A}(\rho^{ABC})$ may then be chosen as
\begin{eqnarray}
\tau^{ABCD} &=& U_{AD} (\rho^{ABC} \otimes \ket{\alpha}\bra{\alpha}_{D}) U_{AD}^{\dagger},\label{dilatare}\\
U_{AD} &=& \sum_{k} U_{A}^{(k)} \otimes  \ket{k}\bra{k}_{D}, \nonumber \\
\ket{\alpha}_{D} &=& \sum_{k} \sqrt{q_{k}} \ket{k}_{D},
\end{eqnarray}
where $\{\ket{k}_D\}$ is an orthonormal basis on system $D$. We shall now make use of Eqs.~(\ref{eq:defqCOV}) and (\ref{eq:WYcorrelationmatrix}) to calculate the matrix ${M^f_\tau}$ corresponding to $\tau^{ABCD}$, relating it to the matrix $M^f_\rho \equiv M^f$ of $\rho^{ABC}$. We will then show that the maximum singular value of $M^f_\tau$ is smaller than that of $M^f$.

To do so we infer from Eq.~(\ref{dilatare}) that the nonzero eigenvalues of $\tau^{ABCD}$ are the same as those of $\rho^{ABC}$, say $\{p_i\}$, while the associated eigenvectors are
\begin{equation}
\ket{\Phi_{i}}_{ABCD}=U_{AD} \ket{\phi_{i}}_{ABC}\otimes \ket{\alpha}_{D},
\end{equation}
$ \ket{\phi_{i}}_{ABC}$ being the eigenvectors of $\rho^{ABC}$. Using the shorthand $\vec{\sigma}=\{\sigma_1,\sigma_2,\sigma_3\}$ as in the main text, we can then write
\begin{eqnarray}
{M^f_\tau} &=& \frac{f(0)}{2} \!\sum_{ij} c^f(p_i,p_j)(p_i \!-\! p_j)^2 \braket{\Phi_{i}|\vec{\sigma}_{A}|\Phi_{j}}\braket{\Phi_{j}|\vec{\sigma}_{B}^{T}|\Phi_{i}} \nonumber \\
&=& \frac{f(0)}{2} \!\sum_{ij} c^f(p_i,p_j)(p_i \!-\! p_j)^2\braket{\phi_{i}|\sum_{k}q_{k}(U_{A}^{(k)})^{\dagger}\vec{\sigma}_{A} U_{A}^{(k)}|\phi_{j}}\braket{\phi_{j}|\vec{\sigma}_{B}^{T} |\phi_{i}} \label{Mftau},
\end{eqnarray}
where we have used the fact that $U_{AD}\ket{\alpha}_{D} = \sum_{k} \sqrt{p_{k}} U_{A}^{(k)} \ket{k}_{D}$. From the well known correspondence between the special unitary group ${\sf SU}(2)$ and special orthogonal group ${\sf SO}(3)$, it follows that for each $k$ there exists an orthogonal matrix $R_{k}$ such that $(U_{A}^{(k)})^{\dagger}\vec{\sigma}_{A} U_{A}^{(k)} = R_{k} \vec{\sigma}_{A}$. Applying this idea to the last line in Eq.~(\ref{Mftau}) we thus obtain
\begin{eqnarray}
M^f_\tau &=&\frac{f(0)}{2} \!\sum_kq_k\sum_{ij} c^f(p_i,p_j)(p_i \!-\! p_j)^2\braket{\phi_{i}|R_k\vec{\sigma}_{A} |\phi_{j}}\braket{\phi_{j}|\vec{\sigma}_{B}^{T} |\phi_{i}} \nonumber\\
&=& S M^f,
\end{eqnarray}
where $S=\sum_{k} q_{k} R_{k}$ is a real matrix such that $SS^T\le\mathbb I$, since it is a convex combination of orthogonal matrices. Since $M^f$ and $M^f_\tau$ are real matrices, their singular values are found as the square roots of the eigenvalues of $Q=M^f(M^f)^T $ and $Q_\tau=M^f_\tau(M^f_\tau)^T=SQS^T,$ respectively. Let $\boldsymbol v$ be the normalised eigenvector of $Q_\tau$ corresponding to its largest eigenvalue. Then
\begin{equation}
\lambda_{\max}(Q_\tau)=\bm{v}^T Q_\tau\bm{v}=\bm{v}^T SQS^T\bm{v}\le \lambda_{\max}(Q)\underbrace{\|S^T\bm{v}\|^2}_{\le 1}\le \lambda_{\max}(Q),
\end{equation}
where we have used that $\bm{v}$ is normalised and $SS^T\le\mathbb I$. This in turn implies that $s_{\max}(M_\tau^f)\leq s_{\max}(M^f)$, concluding our proof.

The proof can be repeated to show monotonicity under unital channels on qubit $B$ as well. This proves that the quantity
  $\widetilde{{\rm Q}}^{f}(\rho^{AB})$ defined in Eq.~(\ref{eq:defqCOVextended}) is a {\it bona fide} quantifier of two-sided quantum correlations which obeys requirement (Q3) for any state $\rho^{AB}$ of a two-qubit system.

%
%


\bibliographystyle{spphys}
\bibliography{quantumcovariance}

%







\end{document}